\documentclass[notitlepage,prd,amsfonts,noshowpacs,nofootinbib]{revtex4-1}
\usepackage{amsmath,graphicx}

\begin{document}
\title{Non-Gaussianities in multi-field DBI inflation with a waterfall phase transition}
\author{Taichi Kidani\footnote{Taichi.Kidani@port.ac.uk}$\|$}
\author{Kazuya Koyama\footnote{Kazuya.Koyama@port.ac.uk}$\|$}
\author{Shuntaro Mizuno
\footnote{shuntaro.mizuno@th.u-psud.fr}$\natural$,$\flat$}
\affiliation{$\|$ Institute of Cosmology and Gravitation, University of Portsmouth, Portsmouth PO1 3FX, UK.}
\affiliation{$\natural$ Laboratoire de Physique Th\'eorique,
Universit\'e Paris-Sud 11 et CNRS, B\^atiment 210, 91405 Orsay
Cedex, France
}
\affiliation{$\flat$ APC (CNRS-Universit\'e Paris 7),
10 rue Alice Domon et L\'eonie Duquet, 75205 Paris Cedex 13,
France}

\date{\today}
\begin{abstract}
We study multi-field DBI inflation models with a waterfall phase transition. This transition happens for a D3 brane moving in the warped conifold if there is an instability along angular directions. The transition converts the angular perturbations into the curvature perturbation. Thanks to this conversion, multi-field models can evade the stringent constraints that strongly disfavour single field ultra-violet DBI inflation models in string theory. We explicitly demonstrate that our model satisfies current observational constraints on the spectral index and equilateral non-Gaussianity as well as the bound on the tensor to scalar ratio imposed in string theory models. In addition we show that large local type non-Gaussianity is generated together with equilateral non-Gaussianity in this model.
\end{abstract}

\maketitle

\section{Introduction}
The inflationary scenario has been established as a standard model for the very early universe not only because it solves the problems of the standard big bang scenario such as
the horizon problem, flatness problem and monopole problem, but also because it explains the origin of the almost scale-invariant spectrum of primordial curvature perturbations
that seeded the Cosmic Microwave Background (CMB) anisotropies (see, e.g. \cite{Komatsu:2010, Larson:2010}).
However, there are numerous models of inflation that are compatible with the current cosmological observations.
Therefore, more precise observations, such as those from the PLANCK satellite \cite{PLANCK}, will help us distinguish between many possible early universe models.

The origin of inflaton, the scalar field that is responsible for inflation, is not specified in many inflation models. Dirac-Born-Infeld (DBI) inflation \cite{Silverstein:2004, Alishahiha:2008} motivated by string theory identifies the inflaton as scalar fields describing the positions of a D-brane in the higher dimensional space in the effective four-dimensional theory. Although DBI inflation is well motivated and it predicts interesting features such as the speed limiting effect on the velocity
of the scalar fields \cite{Underwood:2008} and large non-Gaussianity \cite{Chen:2006nt}, current observations already give strong constrains on the models. In particular, it was shown that the Ultra-Violet (UV) DBI inflation models where a D3 brane is moving down the warped throat is already ruled out by current measurements of the tensor to scalar ratio, the spectral index and non-Gaussianity when one applies microphysics constraints on the variation of the inflaton field in a string theory set-up \cite{Baumann:2007, Lidsey:2007, Ian:2008, Kobayashi:2007hm, Bean:2008}.

However, in the multi-field DBI models, such stringent constraints can be relaxed. In fact, DBI inflation is naturally a multi-inflation model as there are six extra dimensions which are the radial direction and five angular directions in the internal space. In multi-field models, the trajectory in the field space can have a turn and it converts the entropy perturbations into the curvature perturbation on superhorizon scales \cite{Gordon:2001}. It was shown that if there is a sufficient transfer from the entropy perturbations to the curvature perturbation, the constraints on DBI inflation can be significantly relaxed \cite{Langlois:2008wt, Langlois:2008qf, Langlois:2009, Arroja:2008yy}. However, in order to make definite predictions, it is required to calculate the transfer coefficient explicitly with a concrete multi-field potential.

The potential for the angular directions for a D3 brane in the deformed warped conifold was calculated in Ref.~\cite{Baumann:2007ah, Burgess:2007, Chen:2008} and the impact of angular motion on DBI inflation has been studied in Ref.~\cite{spinflation}. Ref.~\cite{Chen:2010} shows that the angular directions can become unstable in a particular embedding of D7 brane on the warped conifold and the angular instability connects different extreme trajectories. These potentials are calculated assuming that the backreaction of the moving brane is negligible and it cannot be applied to DBI inflation directly. However, it is natural to consider that a similar transition due to the angular instability happens also in DBI inflation. The potential derived in Ref.~\cite{Chen:2010} has a similar feature to the potential in hybrid inflation. The mass of the entropy field is large initially. As the brane moves in the radial direction, the mass becomes lighter. Eventually, it arrives at the point where the entropy field becomes tachyonic. Then the inflaton rolls down to the true vacuum along the entropy direction and moves down in the radial direction along the true vacuum.

In this paper, we analyse a two-field DBI model with a potential which has a similar
feature as those obtained in Ref.~\cite{Chen:2010}. Using this potential, we will
explicitly study predictions for observables such as the spectral index, tensor to scalar ratio and non-Gaussainity and see if we can avoid the stringent constraints that rule out the single field UV DBI inflation models. To simplify the calculations, we consider a potential which has an effective single field inflationary attractor with a constant sound speed \cite{Copeland:2010} before and after the tachyonic instability develops along the angular direction. DBI inflation models are known to generate large equilateral type non-Gaussianities \cite{Alishahiha:2008} from the bispectrum of the quantum fluctuations of the scalar field before the horizon exit.
On the other hand, large local type non-Gaussianities \cite{local} (see \cite{Wands:2010af} for a review and references therein) can be generated on super-horizon scales in multi-field models because of the conversion from
the entropy perturbations to the curvature perturbation. Therefore, 
in general multi-field DBI inflation predicts a combination of the equilateral type and local type non-Gausianities and this feature can be used to distinguish DBI models from other inflationary models \cite{Koyama:2010, Babich:2004, Creminelli:2006}. The
presence of both equilateral and local type non-Gaussianities in multi-field DBI inflation was first pointed out by Ref.~\cite{RenauxPetel:2009sj} (see also Ref.~\cite{Emery:2012}). Ref.~\cite{RenauxPetel:2009sj} considered a model where the conversion of the entropy perturbations into the curvature perturbation happens at the end of inflation. In this paper, we consider the case where the conversion happens during inflation by a waterfall phase transition due to the instability along the angular direction.

This paper is organized as follows. In section~\ref{backgrounddynamics}, we briefly review the background dynamics of the single field DBI model with a constant sound speed. Then, the two-field potential that we consider in this paper is
introduced. We also show numerical results for the background dynamics with this potential. In section~\ref{linearperturbation}, the dynamics of the linear cosmological perturbations are studied. We show the results for the power spectrum of the curvature perturbation, the tensor to scalar ratio and the equilateral non-Gaussianity in this model and demonstrate that it is possible to evade the constraints that rule out the single field UV DBI inflation models. In section~\ref{deltansection}, we compute the local-type non-Gaussianity of the curvature perturbation by using the $\delta$N-formalism.
In section~\ref{summarydiscussions}, we summarise this paper. In Appendix A, we review the decomposition in the field space. In appendix B we explain the numerical method to analyse the linear perturbations using the decomposition and the $\delta$N-formalism.

\section{Background dynamics}
\label{backgrounddynamics}
In this section, we first review the background dynamics in the single field DBI inflation model that has a late time attractor solution with a constant sound speed.
Then, we study the background dynamics in the two-field model with a potential that leads to a waterfall phase transition. 

\subsection{The model}
The Lagrangian for the multi-field DBI inflation is given by
\begin{equation}
 P(X^{IJ},\phi^{I}) = \tilde{P}(\tilde{X},\phi^{I}) = - \frac{1}{f(\phi^{I})} \left(\sqrt{1-2 f(\phi^{I}) \tilde{X}} - 1\right) - V\left(\phi^{I}\right),
\label{multifieldaction}
\end{equation}
where $\phi^{I}$ are the scalar fields $(I=1,2,...)$, $f(\phi^I)$ and $V(\phi^I)$ are functions of the scalar fields and $\tilde{X}$ is defined in terms of the determinant \begin{eqnarray}
\mathcal{D} &=& \mbox{det} (\delta^{I}_{J} - 2 f X^{I}_{J} )\nonumber \\
&=& 1 - 2 f G_{IJ} X^{IJ} + 4 f^{2} X^{[I}_{I} X^{J]}_{J} - 8 f^{3} X^{[I}_{I} X^{J}_{J} X^{K]}_{K} + 16 f^{4} X^{[I}_{I} X^{J}_{J} X^{K}_{K} X^{L]}_{L}\, ,
\label{determinant}
\end{eqnarray}
where the brackets denote anti-symmetrisation on the field indices,
\begin{equation}
 \tilde{X} = \frac{(1-\mathcal{D})}{2 f},
\end{equation}
where
\begin{equation}
 X^{IJ} \equiv - \frac{1}{2} \partial _{\mu} \phi^{I} \partial ^{\mu} \phi^{J},
\end{equation}
and $G^{IJ}$ is the metric in the field space. Note that $f(\phi^{I})$ is defined
by the warp factor  $h(\phi^{I})$ and the brane tension $T_3$ as
\begin{equation}
 f(\phi^{I}) \equiv \frac{h(\phi^{I})}{T_{3}}.
\end{equation}
The sound speed is defined as
\begin{equation}
 c_{s} \equiv \sqrt{\frac{\tilde{P}_{,\tilde{X}}}{\tilde{P}_{,\tilde{X}} + 2 \tilde{X} \tilde{P}_{,\tilde{X} \tilde{X}}}} = \sqrt{1- 2 f \tilde{X}},
\end{equation}
where $,_{\tilde{X}}$ means the partial derivative with respect to $\tilde{X}$. Note that $\tilde{X}$ coincides with $X \equiv G_{IJ} X^{IJ}$ in the homogeneous background because all the spatial derivatives vanish. From the action (\ref{multifieldaction}),
we can show that
\begin{equation}
 \tilde{P}_{,\tilde{X}} = \frac{1}{c_{s}}.
\end{equation}
In this paper, we consider the Einstein-Hilbert action for gravity and hence all equations of motion are derived from the action
\begin{equation}
 S = \frac{1}{2} \int d^{4}x \left[^{(4)}R + 2 P(X^{IJ},\phi^{I}) \right],
\label{actionofdbi}
\end{equation}
where we set $8 \pi G = 1$ and $^{(4)}R$ is the four dimensional Ricci curvature.

\subsection{DBI inflation with a constant sound speed}
\label{constantsoundspeed}
Let us consider single field DBI inflation with the potential $V_1(\phi)$
and warp factor $f(\phi)$ where $\phi$ is the scalar field.
Then, the field equation is given by
\begin{equation}
\ddot{\phi} + 3 H \dot{\phi} - \frac{\dot{c_{s}}}{c_{s}} \dot{\phi} + c_{s} 
V_{1,\phi} - \frac{(1-c_{s})^{2}}{2} \frac{f_{,\phi}}{f^{2}} = 0.
\label{single_fieldeq}
\end{equation}

In \cite{Copeland:2010}, it was shown that when $V_1 (\phi)$ and $f(\phi)$
are given by
\begin{equation}
 V_{1}(\phi) = V_{0} \phi^{-q},\label{potentialmizuno}
\end{equation}
\begin{equation}
 f(\phi) = f_{0} \phi^{q+2},\label{warpfactormizuno}
\end{equation}
with constants $V_{0}$, $f_{0}$ and $q$,  
Eq.~(\ref{single_fieldeq})
has a late time attractor inflationary solution with a constant sound speed 
that is given by 
\begin{equation}
 c_{s} = \sqrt{\frac{3}{16 f_{0} V_{0} + 3}}. \label{attractorsoundspeed}
\end{equation}
Throughout this paper, as a concrete example, we consider the case with $q=4$.
This attractor solution is potential dominated, which means that the potential term is much larger than the kinetic term along the attractor solution. If we define the slow-roll parameters as
\begin{equation}
 \epsilon = - \frac{\dot{H}}{H^{2}}, \:\:\:\:\: \eta = \frac{\dot{\epsilon}}{H \epsilon}, \:\:\:\:\: s = \frac{\dot{c_{s}}}{H c_{s}}.
\label{slowrollparameters}
\end{equation}
$s$ vanishes for this attractor solution and also $\epsilon$ and $\eta$ are much less than unity because it is a potential dominated attractor solution.

\subsection{Two-field model}
\label{twofield}
We investigate the following two field model. We define the field $\phi$ as a radial direction and define the field $\chi$ as an angular direction in the warped throat.
The field space metric is then given by
\begin{equation}
 G_{IJ} = A_{I} \delta_{IJ},
\end{equation}
where $A_{\phi} = 1$, $A_{\chi} = \phi^{2}$ and $\delta_{IJ}$ is the kronecker delta.
With this field space metric, 
X becomes
\begin{equation}
 X = \frac{1}{2} \left(\dot{\phi}^{2} + \phi^{2} \dot{\chi}^{2} \right),
 \end{equation}
in the homogeneous background. 
The Friedmann equation is given by 
\begin{equation}
 3 H^{2} = \frac{1}{f(\phi)} \left(\frac{1}{c_{s}} - 1 \right) + V(\phi_{I}).\label{friedmann}
\end{equation}
By varying the action (\ref{actionofdbi}) with respect to the fields, we obtain the equations of motion for the fields as
\begin{equation}
 \ddot{\phi} + 3 H \dot{\phi} - \frac{\dot{c_{s}}}{c_{s}}\dot{\phi} - \phi \dot{\chi}^2 + c_{s} V_{,\phi} - \frac{(1 - c_{s})^2}{2} \frac{f_{,\phi}}{f^2} = 0,\label{phibackgroundeq}
\end{equation}
\begin{equation}
\phi^2 \left(\ddot{\chi} + 3 H \dot{\chi} - \frac{\dot{c_{s}}}{c_{s}}\dot{\chi} \right) + 2 \phi \dot{\phi} \dot{\chi} + c_{s} V_{,\chi} = 0.\label{chibackgroundeq}
\end{equation}
If we specify the potential and the warp factor, the background dynamics are determined by solving Eqs.~(\ref{phibackgroundeq}) and (\ref{chibackgroundeq}). 

In DBI inflation, the scalar fields describe the positions of a brane in the bulk. The explicit form of the multi-field potential depends on the details of the geometry of the warped conifold and various effects from the stabilisation of moduli fields. In Ref.~\cite{Chen:2010}, an example of the multi-field potential that has a similar feature with hybrid inflation was obtained.
In this potential, the inflaton rolls down along the radial direction first. Eventually, it arrives at the transition point where the entropy field becomes tachyonic. Then the inflaton rolls down to the true vacuum along the entropy direction and moves down in the radial direction along the true vacuum.

In this paper, we investigate a two field potential which captures the essential feature of the potential derived in string theory as described above.
We assume the radial field has the form of the potential with a constant sound speed as is discussed in the subsection \ref{constantsoundspeed} to simplify the calculation. The two field potential is given by
\begin{equation}
 V(\phi, \chi)
= \frac{1}{2}\lambda(\chi^2 - \chi^{2}_{0})^2 + g\left(\frac{\chi}{\phi} \right)^2 + \frac{V_{0}}{\phi^{4}}\,,
\label{waterfallpotential}
\end{equation}
Let us assume that the inflaton starts rolling down in the radial direction $\phi$ with a small deviation from $\chi = 0$. Then, in the early stage when $\chi \ll 1$, the 
potential is effectively a single field potential for $\phi$;
\begin{equation}
 V(\phi, \chi) \sim \frac{1}{2}\lambda \chi^{4}_{0} +  \frac{V_{0}}{\phi^{4}}.
\end{equation}
At this stage, if we assume
\begin{equation}
 \frac{1}{2}\lambda \chi^{4}_{0} \ll \frac{V_{0}}{\phi^{4}},\label{conditionichi}
\end{equation}
the effective potential becomes
\begin{equation}
 V_{\rm{eff},1} \sim \frac{V_{0}}{\phi^{4}}.
\end{equation}
\begin{figure}[t]
\centering
\includegraphics[keepaspectratio=true,height=7cm]{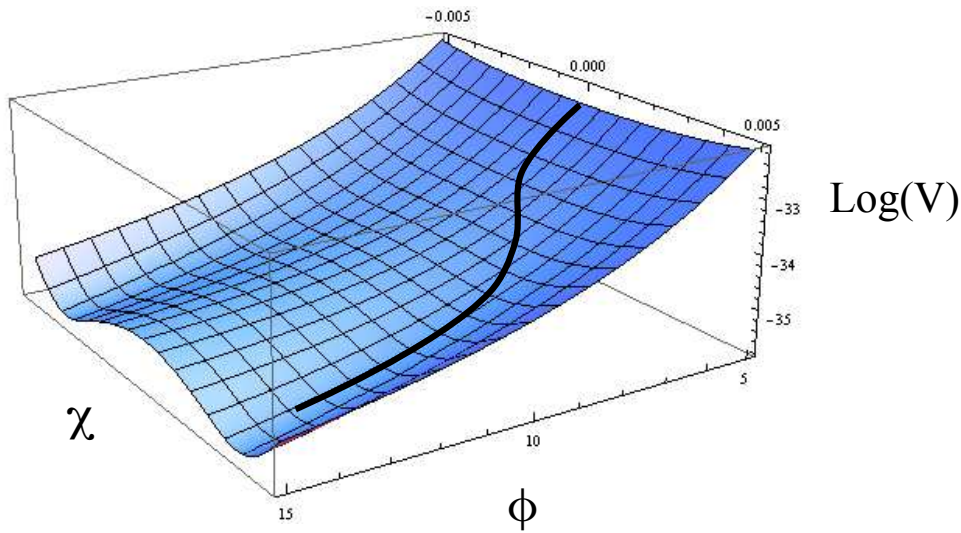}
\caption[Water fall potential]{The potential given by Eq.~(\ref{waterfallpotential}). In order to show the feature of the potential clearly, we plot $\log V(\phi, \chi)$. The parameters are chosen as $\lambda = 3.75 \times 10^{-6}$, $\chi_{0} = 0.004$, $V_{0} = 5 \times 10^{-12}$ and $g = 3 \times 10^{-9}$. For small $\phi$ there is a minimum at $\chi=0$ but for large $\phi$ true vacua appear and the field rolls down to the true vacua causing the waterfall phase transition.}\label{waterfall}
\end{figure}
Because this is in the same form as the potential (\ref{potentialmizuno}), there is a late-time attractor solution with a constant sound speed given by Eq.~(\ref{attractorsoundspeed}).
As the inflaton rolls down in the radial direction, there appears a waterfall phase transition as is shown in Fig.~\ref{waterfall} where the inflaton rolls down to the true minimum of the potential (\ref{waterfallpotential}). This can be seen more clearly by rewriting the potential (\ref{waterfallpotential}) as
\begin{equation}
 V(\phi, \chi) = \frac{1}{2}\lambda \left[\chi^2 - \left(\chi^{2}_{0} - \frac{g}{\lambda \phi^{2}}\right) \right]^2 + V_{0}\left(1 - \frac{g^{2}}{2 \lambda V_{0}} \right)\frac{1}{\phi^{4}} + g\left(\frac{\chi_{0}}{\phi} \right)^2.\label{transformedpotential}
\end{equation}
In this form, $\chi$ appears only in the first term. We can clearly see that $\chi = 0$ is the minimum in the $\chi$ direction when $\phi^{2} < g / (\lambda \chi^{2}_{0})$, while $\chi^{2} = \chi^{2}_{0} - g/(\lambda \phi^{2})$ becomes the minimum in the $\chi$ direction when $\phi^{2} > g / (\lambda \chi^{2}_{0})$. Therefore, $\phi^{2} = g / (\lambda \chi^{2}_{0})$ is the critical transition value for $\phi$. The effective potential in the true vacuum with $\chi^{2} = \chi^{2}_{0} - g/(\lambda \phi^{2})$ is given by 
\begin{equation}
 V(\phi, \chi) \sim V_{0}\left(1 - \frac{g^{2}}{2 \lambda V_{0}} \right)\frac{1}{\phi^{4}} + g\left(\frac{\chi_{0}}{\phi} \right)^2.
\end{equation}
Thus if we assume
\begin{equation}
 \frac{g\left(\chi_{0} / \phi \right)^2}{V_{0}\left(1 - g^{2} / 2 \lambda V_{0} \right) 1/ \phi^{4}} = \frac{g \chi^{2}_{0}}{V_{0}}\frac{\phi^{2}}{1- g^{2} / 2\lambda V_{0}} \ll 1, \label{conditionni}
\end{equation}
the effective potential in the true vacuum becomes
\begin{equation}
 V_{\rm{eff},2} \sim V_{0}\left(1 - \frac{g^{2}}{2 \lambda V_{0}} \right)\frac{1}{\phi^{4}}.
\end{equation}
Again, this is in the form $\sim \phi^{-4}$ and we have a late-time attractor with a constant sound speed
\begin{equation}
 c_{\rm{s}} = \sqrt{\frac{3}{16f_{0}\tilde{V}_{0}+3}},
\end{equation}
where
\begin{equation}
 \tilde{V}_{0} = V_{0}\left(1 - \frac{g^{2}}{2 \lambda V_{0}} \right).
\end{equation}

Now we show our numerical results for the background dynamics. 
We choose the parameters as follows; $\lambda = 3.75 \times 10^{-6}$, $\chi_{0} = 0.004$, $V_{0} = 5 \times 10^{-12}$ and $g = 3 \times 10^{-9}$. The warp factor is given by Eq.(\ref{warpfactormizuno}) with $f_{0} = 1.2 \times 10^{15}$. There 
parameters are chosen so that all the observables satisfy the current 
observational constraints. 

\begin{figure}[h]
\centering
\includegraphics[width=18cm]{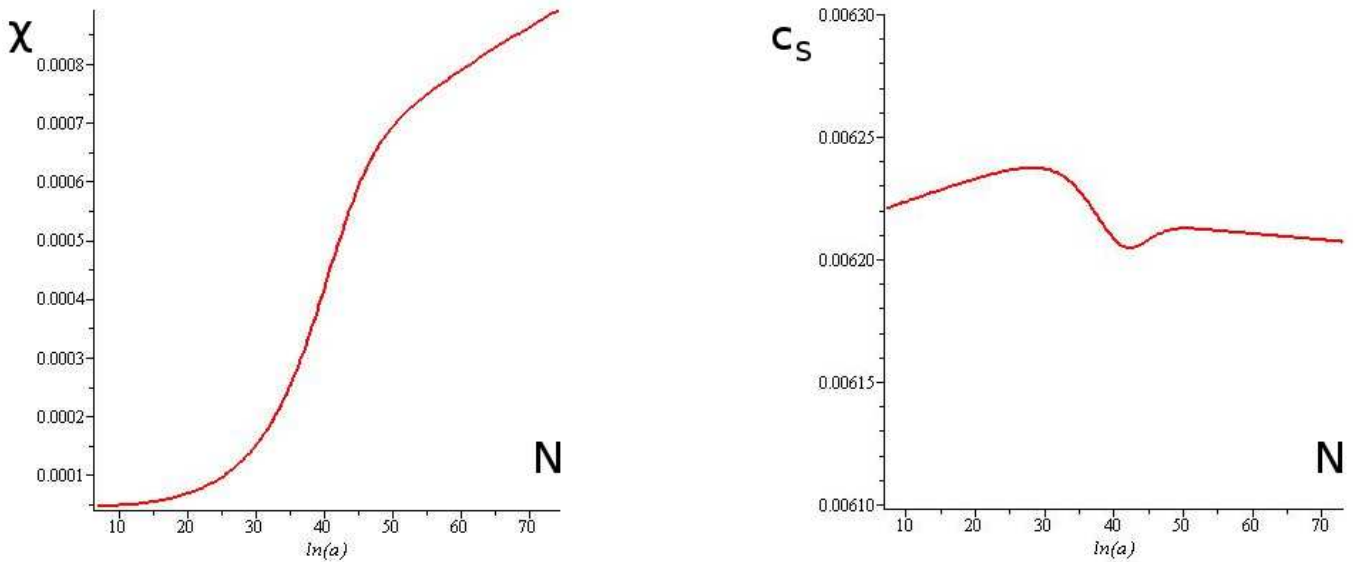}
\caption[The $\chi$ field and the sound speed]{Left: The dynamics of the $\chi$ field. In the early stage ($10 \lesssim N \lesssim 15$), the inflaton rolls down the potential almost in the $\phi$ direction with $\chi \sim 0$. The transition occurs during $20 \lesssim N \lesssim 55$. After $N \sim55$, the inflaton rolls along the true vacuum. Right: The sound speed. The sound speed is almost constant because
it changes very slowly even during the transition.}\label{chisound}
\end{figure}

Firstly, the left panel of Fig.~\ref{chisound} shows the dynamics of the inflaton in the $\chi$ direction. Before the transition happens, the potential has its minimum at $\chi = 0$ in the $\chi$ direction. Therefore, regardless of the initial conditions, the inflaton rolls down the potential and the value of $\chi$ approaches 0 unless the transition occurs while $\chi$ is still large. We use the e-folding number $N=\ln a$
as time. We normalise the e-folding number so that the transition finishes after $N=60$. In this example, $\chi$ is sufficiently small at $N = 10$ and we have an effective single field dynamics until around $N \sim 15$.
As we expected, the inflaton rolls down to the true vacuum in the transition, which occurs during $20 \lesssim N \lesssim 55$. After the transition ends, it rolls down along the true vacuum. Notice that the true vacuum is not along a constant $\chi$ line but the value of $\chi$ along the true vacuum $\chi_{true}$ is a function 
of $\phi$;
\begin{equation}
\chi_{true} = \sqrt{\chi^{2}_{0} - \frac{g}{\lambda \phi^{2}}},
\end{equation}
which can be obtained from Eq.~(\ref{transformedpotential}).
However, as we see in the right panel of Fig.~\ref{chisound}, the trajectory along the true vacuum curves slowly so that the coupling between the adiabatic mode and the entropic mode can be ignored.
Actually, using the value of the Hubble parameter in this model which is $H \sim 2.8 \times 10^{-8}$, we can roughly estimate how many e-folds we need after sound horizon exit of the mode which we are considering \cite{Liddle:2009} assuming 
instant reheating. In our model, it is around 60 e-folds. We can see that the transition ends within 60 e-folds after sound horizon exit if we consider modes that exit the sound horizon in the effective single field regime ($10 < N < 15$). We 
assume inflation ends after the transition by some mechanisms such as an annihilation
of D brane with anti-D brane. 

\begin{figure}[h]
\centering
\includegraphics[width=18cm]{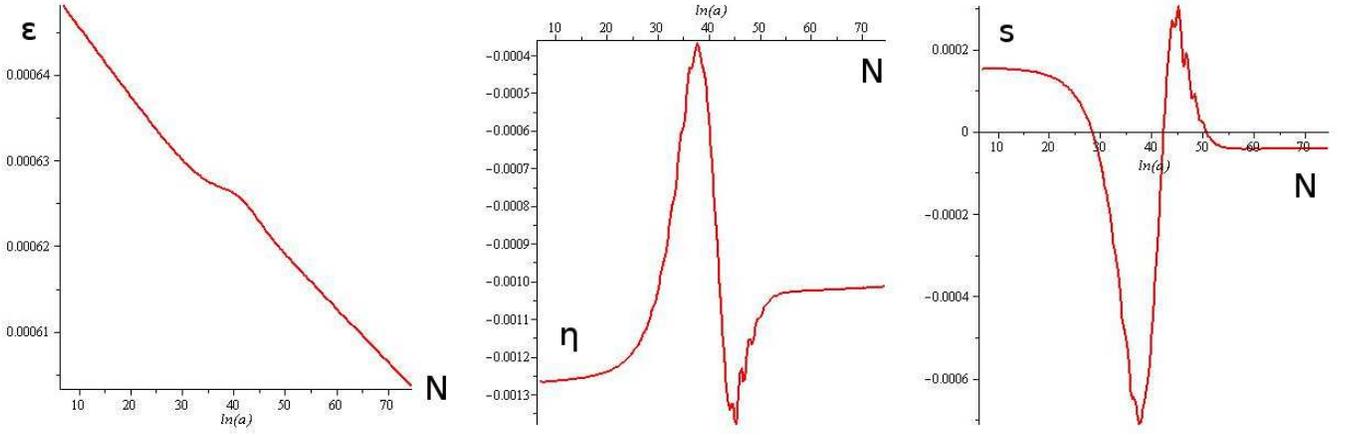}
\caption[Slow-roll parameters]{Left: The slow-roll parameter $\epsilon$. It is always smaller than $1 \times 10^{-3}$ even during the transition.
Middle: The slow-roll parameter $\eta$. Its absolute value is always smaller than $2 \times 10^{-3}$.
Right: The slow-roll parameter $s$. $\lvert s \rvert$ takes the largest value $-7 \times 10^{-4}$ during the transition. However, its absolute value is still much less than one.}\label{slowroll}
\end{figure}

The right panel of Fig.~\ref{chisound} shows the sound speed. Before the transition, the sound speed slowly changes. This is because the condition (\ref{conditionichi}) is not fully satisfied. On the other hand, after the transition, we can clearly see that the sound speed is almost constant. The sound speed changes the most during the transition and the slow-roll parameter $s$ takes the largest value during the transition. However, the largest value of $s$ is still around $-7 \times 10^{-4}$. 
Actually, as is shown in Fig.~\ref{slowroll}, all the slow-roll parameters are always much smaller than unity even during the transition. Therefore, the slow-roll approximation always holds in this model and the sound speed is almost constant even during the transition.

\section{Linear perturbation}
\label{linearperturbation}
In this section, we study linear perturbations in the two field DBI model introduced in section~\ref{backgrounddynamics}. We first derive the coupled equations for 
adiabatic and entropy perturbations. Secondly, we present the numerical results for the power spectrum of the curvature perturbation.
Then, we show that the constraint on the tensor to scalar ratio in the single field DBI inflation is incompatible with current observations and show how this constraint can be relaxed in the multi-field models. Finally, we demonstrate that the model
considered in this paper can actually evade the constraints. 

\subsection{Field perturbations}
We introduce the linear perturbations of the fields $Q^{\phi}$ and $Q^{\chi}$ as
\begin{equation}
 \phi=\phi_{0}(t) + Q^{\phi}\left(x,t\right), \: \: \: \: \: \chi=\chi_{0}(t) + Q^{\chi}\left(x,t\right),
\end{equation}
As discussed in appendix~\ref{Adiabaticandentropic}, we decompose these perturbations into the instantaneous adiabatic and entropy perturbations as
\begin{equation}
 Q^{\phi}=Q_{\sigma}e_{\sigma}^{\phi} + Q_{s}e_{s}^{\phi},
\end{equation}
\begin{equation}
 Q^{\chi}=Q_{\sigma}e_{\sigma}^{\chi} + Q_{s}e_{s}^{\chi},
\end{equation}
where $e^{I}_{\sigma}$ and $e^{I}_{s}$ are the adiabatic and entropic basis, respectively. They are defined as 
\begin{equation}
 e_{\sigma}^{\phi}=\sqrt{c_{s}}\frac{\dot{\phi}}{\sqrt{\dot{\phi}^{2}+\phi^{2}\dot{\chi}^{2}}}, \: \: \: \: \: e_{s}^{\phi}=\frac{1}{\sqrt{c_{s}}}\frac{\phi \dot{\chi}}{\sqrt{\dot{\phi}^{2}+\phi^{2}\dot{\chi}^{2}}},
\label{basisone}
\end{equation}
\begin{equation}
 e_{\sigma}^{\chi}=\sqrt{c_{s}}\frac{\dot{\chi}}{\sqrt{\dot{\phi}^{2}+\phi^{2}\dot{\chi}^{2}}}, \: \: \: \: \: e_{s}^{\chi}=\frac{1}{\sqrt{c_{s}}}\frac{-\dot{\phi}}{\phi\sqrt{\dot{\phi}^{2}+\phi^{2}\dot{\chi}^{2}}}.
\label{basistwo}
\end{equation}
For the analysis of perturbations, it is convenient to use the conformal time $\tau = \int dt/a(t)$ and define the canonically normalized fields as
\begin{equation}
 v_{\sigma}=\frac{a}{c_{s}} Q_{\sigma}, \: \: \: \: \: v_{s}=\frac{a}{c_{s}} Q_{s}.
\label{relationofvandq}
\end{equation}
The equations of motion for $v_{\sigma}$ and $v_{s}$ are obtained as 
\begin{equation}
 v_{\sigma}'' - \xi v_{s}' + \left(c_{s}^{2} k^{2} - \frac{z''}{z} \right)v_{\sigma} - \frac{\left(z \xi \right)'}{z} v_{s} = 0,\label{equationofmotionone}
\end{equation}
\begin{equation}
 v_{s}'' + \xi v_{\sigma}' + \left(c_{s}^{2} k^{2} - \frac{\alpha''}{\alpha} + a^{2} \mu_{s}^{2} \right)v_{s} - \frac{z'}{z} \xi v_{\sigma} = 0,\label{equationofmotiontwo}
\end{equation}
where the prime denotes the derivative with respect to $\tau$ and
\begin{equation}
 \xi \equiv \frac{a}{\dot{\sigma}} \left[\left(1+c_{s}^{2}\right) \tilde{P}_{,s} - c_{s}^{2} \dot{\sigma}^{2} \tilde{P}_{,\tilde{X} s} \right],
\end{equation}
\begin{equation}
 \mu_{s}^{2} \equiv -c_{s} \tilde{P}_{,ss} - \frac{1}{\dot{\sigma}^{2}} \tilde{P}_{,s}^{2} + 2 c_{2}^{2} \tilde{P}_{,\tilde{X} s} \tilde{P}_{,s},
\end{equation}
\begin{equation}
 z \equiv \frac{a \dot{\sigma}}{\sqrt{c_{s}} H}, \: \: \: \: \: \alpha \equiv a \frac{1}{\sqrt{c_{s}}},
\end{equation}
with
\begin{equation}
 \dot{\sigma} \equiv \sqrt{2 X}, \:\:\: \tilde{P}_{s} \equiv \tilde{P}_{,I} e_{s}^{I} \sqrt{c_{s}}, \:\:\: \tilde{P}_{,\tilde{X} s} \equiv \tilde{P}_{,\tilde{X} I} e_{s}^{I} \sqrt{c_{s}}, \:\:\: \tilde{P}_{,ss} \equiv \left(\mathcal{D}_{I} \mathcal{D}_{J} \tilde{P} \right) e_{s}^{I} e_{s}^{J} c_{s},
\end{equation}
where $\mathcal{D}_{I}$ denotes the covariant derivative with respect to the field space metric $G_{IJ}$. We numerically solve Eqs.~(\ref{equationofmotionone}) and (\ref{equationofmotiontwo}) to compute the curvature perturbation. 

\subsection{Curvature perturbation}
\label{subcurvature}
\begin{figure}[h]
\centering
\includegraphics[width=18cm]{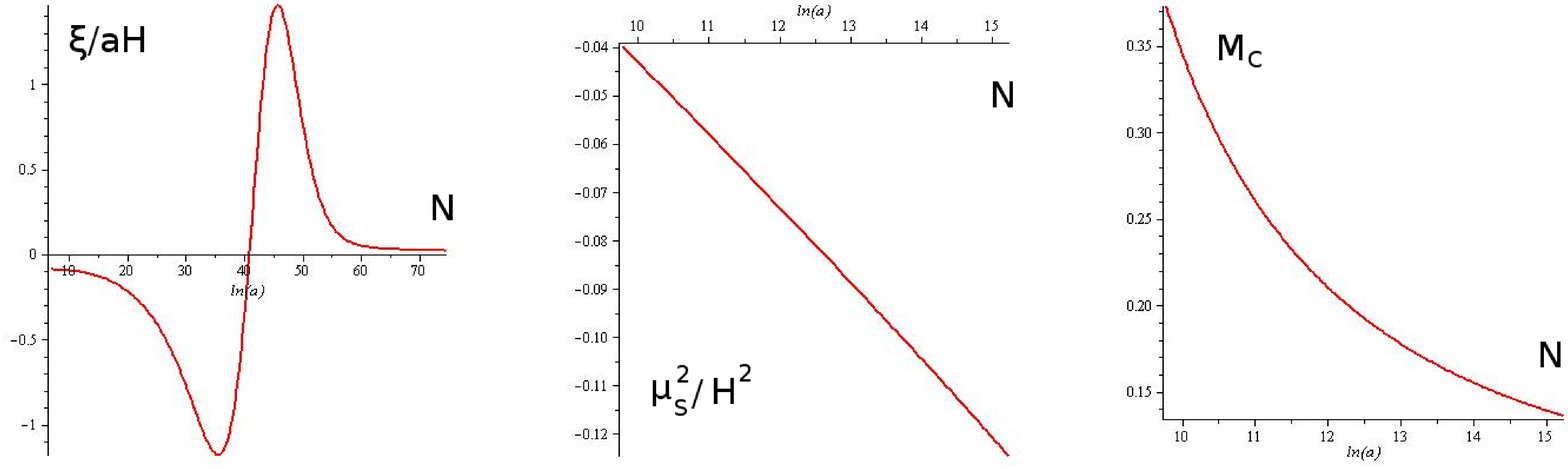}
\caption[The mass and its change]{Left: The coupling $\xi/aH$. Before and after the transition the coupling is small but it becomes large during the transition converting the entropy perturbation into the curvature perturbation. Middle: The mass of the entropy perturbation divided by the Hubble parameter squared. Its absolute value is much smaller than one around sound horizon exit at $N \sim 10$.
Right: The quantity $M_{c}$ that shows how the mass of the entropy perturbation changes. It is also smaller than one around sound horizon exit.}\label{massofentropy}
\end{figure}

In this subsection, we show how we set the initial conditions for Eqs.~(\ref{equationofmotionone}) and (\ref{equationofmotiontwo}) to calculate the 
power spectrum of the curvature perturbation. If the trajectory is not curved significantly, the coupling $\xi/a H$ becomes much smaller than one. 
In Fig.~\ref{massofentropy}, we see that $\xi/a H$ is still much smaller than one before the transition starts around $N \sim 20$. Also, the slow-roll conditions are satisfied as is shown in section~\ref{twofield} and this means that H, $\dot{\sigma}$ and $c_{s}$ change very slowly with time compared to the Hubble scale so that the approximations $z''/z \simeq 2 / \tau^{2}$ and $\alpha'' / \alpha \simeq 2 / \tau^{2}$ hold. Therefore, we can approximate the Eqs.~(\ref{equationofmotionone})
and ~(\ref{equationofmotiontwo}) as Bessel differential equations before $N \sim 20$ (see appendix~\ref{numericalmethod} for the entropy perturbation). Then, the solutions with the Bunch-Davis vacuum initial conditions are given by
\begin{equation}
 v_{\sigma k} \simeq \frac{1}{\sqrt{2 k c_{s}}} e^{-i k c_{s} \tau} \left(1 - \frac{i}{k c_{s} \tau} \right),\label{solone}
\end{equation}
\begin{equation}
 v_{s k} \simeq \frac{1}{\sqrt{2 k c_{s}}} e^{-i k c_{s} \tau} \left(1 - \frac{i}{k c_{s} \tau} \right),\label{soltwo}
\end{equation}
when $\mu^{2}_{s} / H^{2}$ is negligible for the entropy mode. Then, the power spectra of $Q_{\sigma}$ and $Q_{s}$ are obtained as
\begin{equation}
 P_{Q_{\sigma}} \simeq \frac{H^{2}}{4 \pi^{2} c_{s}}, \:\:\:\:\: P_{Q_{s}} \simeq \frac{H^{2}}{4 \pi^{2} c_{s}},
\label{pqsigmas}
\end{equation}
which are evaluated at sound horizon crossing. The power spectrum of $\mathcal{R}$ at sound horizon crossing is given by
\begin{equation}
 P_{\mathcal{R}_{*}} = \left. \frac{k^{3}}{2 \pi^{2}} \lvert \mathcal{R} \rvert^{2} \right\rvert_{*}  = \left. \frac{k^{3}}{2 \pi^{2}} \frac{\lvert v_{\sigma k} \rvert^{2}}{z^{2}} \right\rvert_{*} \simeq \left. \frac{H^{4}}{4 \pi^{2} \dot{\sigma}^{2}} \right\rvert_{*} = \left. \frac{H^{2}}{8 \pi^{2} \epsilon c_{s}} \right\rvert_{*},
\label{curvaturepowerspectrum}
\end{equation}
where the subscript $*$ indicates that the corresponding quantity is evaluated at sound horizon crossing $k c_{s} = a H$. \\

We investigate a mode which exits the sound horizon at $N \sim 10$ where we have an effectively single field dynamics. As stated above, the coupling $\xi/aH$ is much smaller than unity around sound horizon exit. Also, as we can see in Fig.~\ref{massofentropy}, $\lvert \mu^{2}_{s} / H^{2} \rvert$ is also much smaller than unity around sound horizon exit at $N \sim 10$. The mass also changes very slowly. If we define a quantity
\begin{equation}
 M_{c} \equiv \frac{\dot{\mu}_{s}}{\mu_{s} H},
\end{equation}
which quantifies how rapidly the mass of the entropy perturbation changes, we can see in Fig.~\ref{massofentropy} that $M_{c}$ is smaller than unity for at least 5 e-folds after sound horizon exit. Therefore, we can set the initial conditions for Eqs.~(\ref{equationofmotionone}) and (\ref{equationofmotiontwo}) by the solutions (\ref{solone}) and (\ref{soltwo}). Note that we set the initial conditions at $N \sim 7$ when the mode which we consider is still well within the sound horizon.

We treat $v_{\sigma}$ and $v_{s}$ as two independent stochastic variables for the modes well inside the sound horizon as in \cite{Tsujikawa:2003}.
This means that we perform two numerical computations to obtain $P_{\mathcal{R}}$ for example. One computation corresponds to the Bunch Davis vacuum state
for $v_{\sigma}$ and $v_{s}$ is set to be zero to obtain the solution $\mathcal{R}_{1}$. Another computation corresponds to
the Bunch Davies vacuum state for $v_{s}$ and $v_{\sigma}$ is set to be zero, in which case we obtain the solution $\mathcal{R}_{2}$. Then, the curvature power spectrum can be expressed as a sum of two solutions;
\begin{equation}
 P_{\mathcal{R}} = \frac{k^{3}}{2 \pi^{2}} \left(\lvert \mathcal{R}_{1} \rvert^{2} + \lvert \mathcal{R}_{2} \rvert^{2} \right).
\end{equation}
This procedure is applied to all the numerical computations in this paper.

\begin{figure}[h]
\centering
\includegraphics[width=18cm]{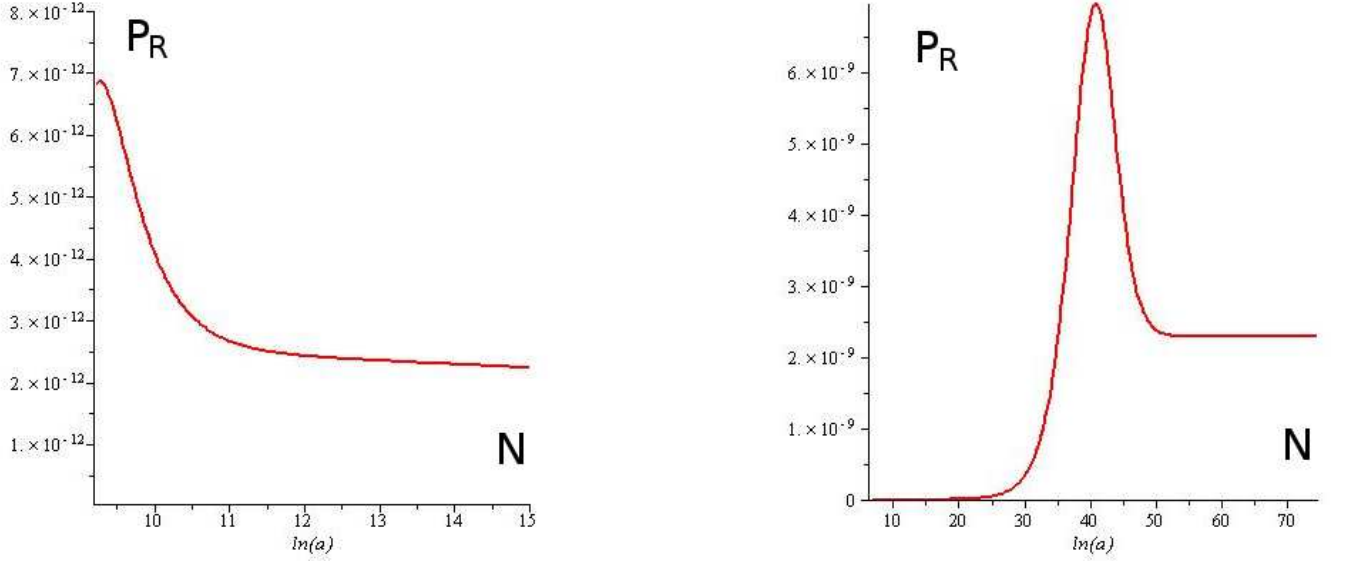}
\caption[The curvature power spectrum]{Left: The power spectrum of the curvature perturbation becomes constant a few e-folds after sound horizon exit which is around $N \sim 17$.
Right: We can see that the power spectrum of the curvature perturbation is enhanced during the transition.
Then, it becomes constant again after the transition. Note that both of these figures describe the contribution from the real part of $v_{\sigma k}$ and the contribution
from the imaginary part shows the same behaviour. Therefore, it is sufficient to show only these figures in order to know its behaviour
because the power spectrum of the curvature perturbation is just a sum of those two contributions.}\label{curvature}
\end{figure}

Fig.~\ref{curvature} shows the numerical results for the power spectrum of the
curvature perturbation. As we can see from the small value of $\epsilon$, the Hubble parameter $H$ changes very slowly. It changes only by a few percent between $N \sim 10$ and $N \sim 70$. Substituting $H \sim 2.8 \times 10^{-8}$, $c_{\rm{s}} \sim 6.2 \times 10^{-3}$ and $\epsilon = 6.4 \times 10^{-4}$ into 
Eq.~(\ref{curvaturepowerspectrum}), we obtain the value of the curvature power spectrum after sound horizon exit as
\begin{equation}
 P_{\mathcal{R}_{*}} = \left. \frac{H^{2}}{8 \pi^{2} \epsilon c_{s}} \right\rvert_{*} \sim 2.5 \times 10^{-12}.
\end{equation}
This should coincide with the result of the numerical solution.
Actually, in the left figure of Fig.~\ref{curvature}, it is shown that $P_{\mathcal{R}}$ becomes almost constant soon after sound horizon exit and it is given by $P_{\mathcal{R}} \sim 2.5 \times 10^{-12}$.
It does not change significantly until $N \sim 15$. When the trajectory in 
field space curves, the curvature perturbation is sourced by
the entropy perturbation and it is enhanced. We can actually see that the power spectrum of the curvature perturbation is enhanced by a factor of $\sim 9 \times 10^{2}$ during the transition in the right figure of Fig.~\ref{curvature}. After the transition, it takes a constant value $P_{\mathcal{R}} \sim 2.3 \times 10^{-9}$, which is compatible with the CMB observation \cite{Komatsu:2010}. 

If we express the final curvature power spectrum as \cite{Langlois:2009}
\begin{equation}
 P_{\mathcal{R}} = \frac{P_{\mathcal{R}_{*}}}{\cos^{2}{\Theta}},
\label{rainy}
\end{equation}
the enhancement is quantified by the function $\cos^{2}{\Theta}$. In this model, we have $\cos^{2}{\Theta} \sim 1.1 \times 10^{-3}$.
When $\lvert \mu^{2}_{s} / H^{2} \rvert$ is much smaller than unity, from Eq. (\ref{rainy}), the spectral index $n_{s}$ is given by
\begin{equation}
 n_{s} - 1 \equiv \frac{d \ln {P_{\mathcal{R}}}}{d \ln {k}} = -2 \epsilon_{*} - \eta_{*} - \alpha_{*} \sin {2 \Theta} - 2 \beta_{*} \sin^{2} {\Theta},
\label{spectralindex}
\end{equation}
where
\begin{equation}
 \alpha = \frac{\xi}{a H},\: \: \beta = \frac{s}{2} - \frac{\eta}{2} - \frac{1}{3H^{2}} \left( \mu^{2}_{s} + \frac{\Xi^{2}}{c^{2}_{s}} \right), \: \: \Xi \equiv \frac{c_{s} \xi}{a},
\end{equation}
where only the leading order terms in the slow-rolling approximation are kept in the expression for $\beta$. Also, the parameter which quantifies the equilateral non-Gaussianity is expressed as
\begin{equation}
 f_{NL}^{equil} = - \frac{35}{108} \frac{1}{c_{s}^{2}} \cos^{2} {\Theta} \sim - \frac{\cos^{2} {\Theta}}{3 c^{2}_{s}}.
\label{fnlequilateral}
\end{equation}

Because the amplitude of the tensor modes are not affected by the scalar field dynamics their power spectrum is given by
\begin{equation}
 P_{\mathcal{T}} = \left. \frac{2 H^{2}}{\pi^{2}} \right\rvert_{*}.
\end{equation}
Therefore, the tensor to scalar ratio is expressed as
\begin{equation}
 r \equiv \frac{P_{\mathcal{T}}}{P_{\mathcal{R}}} = \left. 16 \epsilon c_{s} \right\rvert_{*} \cos^{2} {\Theta}.
\label{tensortoscalar}
\end{equation}

\subsection{Gravitational waves constraints}
\label{gravitationalwaves}
In Ref. \cite{Lidsey:2007}, it is shown that the single field ultra-violet (UV) DBI inflation is disfavoured by observation as follows.
Firstly, the tensor to scalar ratio is related to the field variation $\Delta \phi$ by the Lyth bound
\begin{equation}
 \frac{1}{M_{P}^{2}} \left( \frac{\Delta \phi}{\Delta N} \right)^{2} = \frac{r}{8},
\label{lythbound}
\end{equation}
where $N \equiv \int dt H$. Because the field variation $\Delta \phi$ corresponds to the radial size of the extra dimensions,
it is constrained by the size of the extra dimensions. From Eq. (\ref{lythbound}), the upper bound on $\Delta \phi$ gives the model independent upper bound on the tensor to scalar ratio for standard UV DBI inflation. The bound is typically given by 
\begin{equation}
 r < 10^{-7},
\label{upper}
\end{equation}
when we assume the minimum number of e-foldings that could be probed by observation as $\Delta N \sim 1$.
Secondly, the lower bound on the tensor to scalar ratio in the single field UV DBI inflation is derived in the following way.
The relation (\ref{spectralindex}) can be rewritten as
\begin{equation}
 1 - n_{\rm{s}} \simeq \frac{\sqrt{3 \lvert f^{\rm{equil}}_{\rm{NL}} \rvert}r}{4 \cos^{3}{\Theta}} - \frac{\dot{f}}{Hf} + \alpha_{*} \sin{2 \Theta} + 2 \beta_{*} \sin^{2}{\Theta},
\label{multispectral}
\end{equation}
by using the Eqs.~(\ref{slowrollparameters}) and (\ref{fnlequilateral}) as shown in Ref.~\cite{Langlois:2009}. Note that a term proportional to $c_{s}^{2} s_{*}$ is neglected because both $c_{s}$ and $s_{*}$ are small. For the single field UV DBI inflation, we have $\dot{f} > 0$ and $\Theta = 0$. This gives
\begin{equation}
 r > \frac{4}{\sqrt{3 \lvert f^{\rm{equil}}_{\rm{NL}} \rvert}} \left( 1 - n_{s} \right) \: \: \: \: \: \: \: \: \: \: (\rm{single \:\: field}),
\label{david}
\end{equation}
from Eq. (\ref{david}). The amplitude of the equilateral non-Gaussianity is constrained as 
\begin{equation}
 -151 < f_{NL}^{equil} < 253 \: \: \: \: \: \: \: \: \: \: \: \: \rm{at\:\: 95\% \:\: C.L.},
\end{equation}
from WMAP5 and the best-fit value for the specrtral index is $n_{s} \simeq 0.982$ \: \cite{Larson:2010}. From those values, we can obtain the lower bound on the tensor to scalar ratio as
\begin{equation}
 r \gtrsim 10^{-3}.
\label{lower}
\end{equation}
Clearly, the lower bound (\ref{upper}) is not compatible with the upper bound (\ref{lower}). This is why single field UV DBI inflation is disfavoured by observation.

These constraints are relaxed when we consider the multi-field models. The upper bound is relaxed because we have angular directions and the
field variation is not only determined by the radial coordinate. More importantly, the lower bound is relaxed significantly. In Eq.~(\ref{multispectral}), the last two terms become important if there is a transfer from entropy to adiabatic modes ($\Theta \neq 0$). In the model considered here, the curvature perturbation originated from the 
entropy perturbation dominates the final curvature perturbation and the spectral index is indeed determined by the mass of the entropy mode
\begin{equation}
 n_{s} \sim \left. \frac{2 \mu_s^2}{3 H^2} \right \rvert_{*} + 1 \sim  0.972,
\end{equation}
which is compatible with the WMAP observation. This value has also been confirmed in our numerical computations for the curvature perturbation. Thus there is no longer the lower bound for the tensor to scalar ratio. The tensor to scalar ratio is obtained 
by substituting $\epsilon \sim 6.4 \times 10^{-4}$, $c_{s} \sim 6.2 \times 10^{-3}$ and $\cos^{2} {\Theta} \sim 1.1 \times 10^{-3}$ into Eq. (\ref{tensortoscalar}) as
\begin{equation}
 r = \left. 16 \epsilon c_{s} \right\rvert_{*} \cos^{2} {\Theta} \simeq 7.0 \times 10^{-8}.
\end{equation}
This is compatible with the upper bound (\ref{upper}). 

\section{Non-Gaussianities}
\label{deltansection}
In this section, we calculate non-Gaussianities of the curvature perturbation. In addition to the equilateral type non-Gaussianity, the transition may generate local type non-Gaussianity. The local type non-Gaussianity can be easily calculated by the $\delta N$ formalism. We first briefly review the $\delta$N-formalism \cite{Lyth:2005, Sasaki:1996} and compute the power spectrum of the curvature perturbation in order to confirm the accuracy of the $\delta N$ formalism in our model. Then, we use it to compute the non-Gaussianities of the primordial curvature perturbation.

\subsection{$\delta N$ formalism}
In the $\delta$N-formalism, the curvature perturbation on the uniform density hypersurface $\zeta$ evaluated at some time $t = t_{f}$ is identified as the difference between the number of e-folds $N_{A}(t_{f},t_{i},\mathbf{x})$ and $N_{0}(t_{f},t_{i})$ where $N_{A}(t_{f},t_{i},\mathbf{x})$ is the number of e-folds
from an initial flat slice at $t = t_{i}$ to a final uniform density slice at $t = t_{f}$ and $N_{0}(t_{f},t_{i})$ is the number of e-folds
from an initial flat slice at $t = t_{i}$ to a final flat slice at $t = t_{f}$;
\begin{equation}
\zeta(t_{f}, \mathbf{x}) \simeq \delta N \equiv N_{A}(t_{f},t_{i},\mathbf{x}) - N_{0}(t_{f},t_{i}).
\end{equation}
We can ignore the dependence of $N$ on the time derivatives of the fields when the slow-roll approximations hold because
the equations of motion for perturbations are reduced to first order differential equations. Then, we can expand $\delta N$ in terms of the field values at the 
sound horizon crossing up to the second order
\begin{eqnarray}
 \delta N \left(\phi_{\sigma}, \dot{\phi}_{\sigma}, \phi_{s}, \dot{\phi}_{s} \right) &\simeq& \delta N \left(\phi_{\sigma}, \phi_{s} \right) \\
 &\simeq& N_{,\phi_{\sigma}} Q_{\sigma} + N_{,\phi_{s}} Q_{s} + \frac{1}{2} N_{\phi_{\sigma} \phi_{\sigma}} Q_{\sigma}^{2} + \frac{1}{2} N_{\phi_{s} \phi_{s}} Q_{s}^{2} \, ,\label{ndecomposition}
\end{eqnarray}
where $\phi_{\sigma}$ denotes the scalar field in the instantaneous adiabatic direction which is almost equivalent to $\phi$ before the transition and
$\phi_{s}$ denotes the scalar field in the instantaneous entropic direction which is almost equivalent to $\chi$ before the transition (see Fig.~\ref{chisound}).
Note that we ignore the cross term $Q_{\sigma} Q_{s}$ because the two fields are independent quantum fields around sound horizon exit before the transition.

\begin{figure}[h]
\centering
\includegraphics[width=15cm]{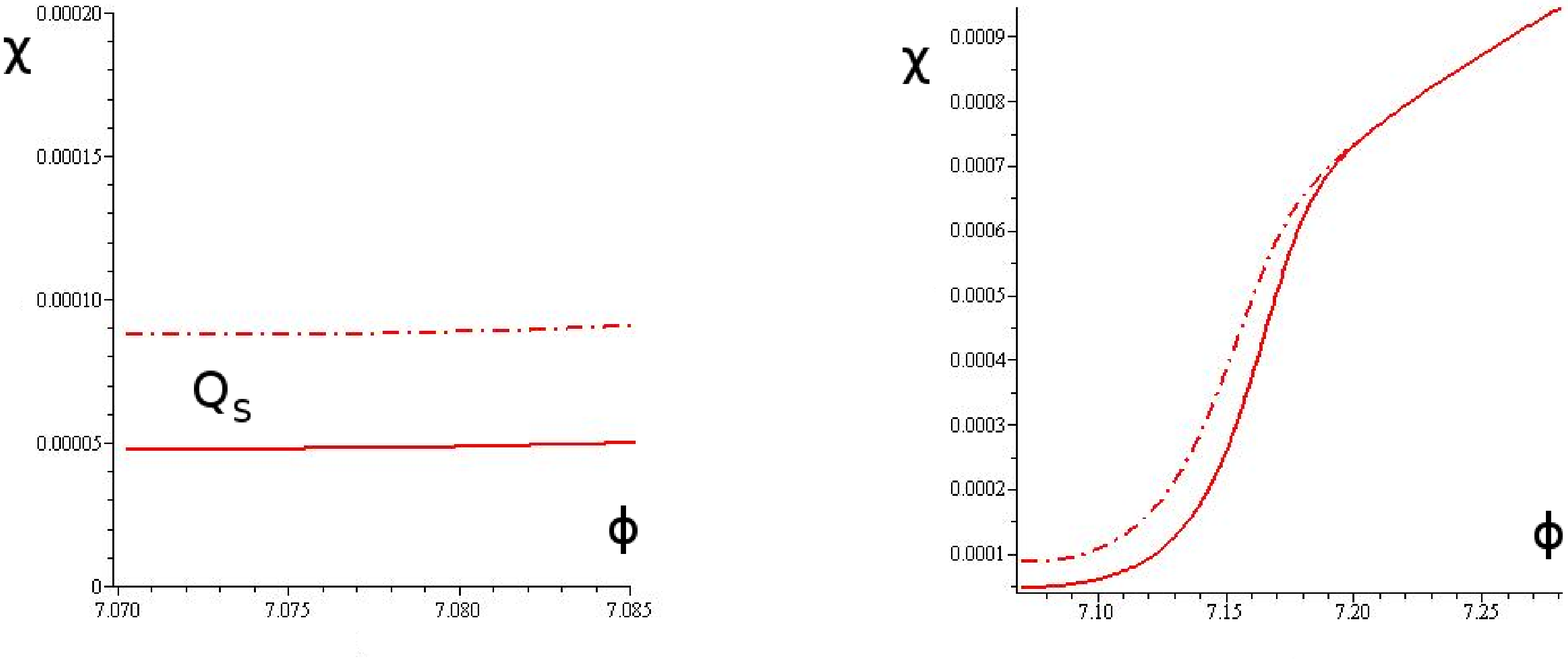}
\caption[$\delta$N in the entropic direction]{Left: Two trajectories
in the field space obtained by perturbing the initial value of the entropy field. 
The solid line describes the unperturbed trajectory
while the dotted line describes the perturbed trajectory.
Right: The transition occurs at different values of $\phi$. Again, the solid line is the unperturbed trajectory
while the dotted line is the perturbed trajectory.}\label{entropyperturbation}
\end{figure}

We compute the power spectrum of the curvature perturbation at $t = t_{f}$ after the transition by the $\delta N$-formalism taking $t_{i}$ to be well before the transition when the curvature power spectrum is constant after sound horizon exit.
In this case, as is shown in Fig.~\ref{entropyperturbation}, the transition occurs at different $\phi_{\sigma}$ if we perturb the initial field values in the entropic direction. Because taking perturbations in the initial field values changes the trajectory unlike single field cases, we need to be careful about the definition of the final slice. Let us define $\phi_{\sigma}^{I}$ as the field values along the unperturbed trajectory (solid line) and $\tilde{\phi}_{\sigma}^{I}$ as the field values along the perturbed trajectory (dotted line) in Fig.~\ref{entropyperturbation}. We define $\delta$N as
\begin{equation}
 \delta N = \int ^{\tilde{t}_{f}}_{\tilde{t}_{i}} \tilde{H} d\tilde{t} - \int ^{t_{f}}_{t_{i}} H dt,
\label{deltanentropy}
\end{equation}
where the tilde denotes the quantities with the perturbed initial conditions.
For example, if we perturb the initial field values to the entropic direction, the perturbed initial field values $\tilde{\phi}_{\sigma}(\tilde{t}_{i})$ corresponds to the position in the field space as
\begin{equation}
 \tilde{\phi}_{\sigma}^{I}(\tilde{t}_{i}) = \phi_{\sigma}^{I}(t_{i}) + Q_{s} e_{s}^{I}, \label{entropicperturbedfield}
\end{equation}
with $e_{s}^{I}$ given in Eqs.~(\ref{basisone}) and (\ref{basistwo}). Note that $\tilde{t}_{f}$ in Eq.~(\ref{deltanentropy}) is the time when $\tilde{\phi}_{\sigma}^{I}$ takes the same value
as $\phi_{\sigma}^{I}(t_{f})$ well after the transition. Because both trajectories merge into the attractor solution in the true vacuum after the transition as is shown in Fig.~\ref{entropyperturbation}, we take the final slice so that both unperturbed and perturbed trajectories have the same field values $\phi_{f}$ and $\chi_{f}$ on the final slices. Because we can determine all the phase space variables
if we know the values of the fields in the slow-roll case, this means that we have the same $\phi$, $\dot{\phi}$, $\chi$ and $\dot{\chi}$ on the final slices which results in the same H (i.e. uniform density) from Eq. (\ref{friedmann}). By using the definition of $\delta$N in Eq.~(\ref{deltanentropy}), we can numerically compute the quantity
\begin{equation}
 N_{,\phi_{s}} = \frac{N\left(\phi_{\sigma}^{I}(t_{i}) + Q_{s} e_{s}^{I}  \right) - N\left(\phi_{\sigma}^{I}(t_{i}) - Q_{s} e_{s}^{I} \right)}{2Q_{s}},
\label{nphis}
\end{equation}
where we make $Q_{\sigma}$ sufficiently small so that the value of $N_{,\phi_{\sigma}}$ does not depend on the value of $Q_{\sigma}$.
Because the contribution from $N_{,\phi_{\sigma}}$ in Eq.(\ref{ndecomposition}) is negligible in our model, we obtain
\begin{equation}
 P_{\mathcal{R}} = P_{\zeta} \simeq N_{,\phi_{s}}^{2} P_{Q_{s}} \rvert _{t=t_{i}},\label{deltanentropic}
\end{equation}
where $P_{Q_{s}}$ is given by Eq.~(\ref{pqsigmas}). Note that the curvature perturbation on the uniform density hypersurface coincides with the comoving
curvature perturbation on super-horizon scales \cite{Bassett:2006}. The numerical result shows that the $\delta$N-formalism successfully
predicts the value of the final curvature perturbation $P_{\mathcal{R}} \sim 2.3 \times 10^{-9}$ within a few percent error in this model.

\subsection{Non-Gaussianities}
Now, we can compute the non-Gaussianities of the curvature perturbation using the $\delta N$ formalism as follows.
The bispectrum of the curvature perturbation is defined as
\begin{equation}
 \left< \zeta(\vec{k_1}) \zeta(\vec{k_2}) \zeta(\vec{k_3}) \right> = \left( 2 \pi \right)^{2} B_{\zeta}(k_1, k_2, k_3) \delta^{(3)} (\vec{k_1}+\vec{k_2}+\vec{k_3}).
\label{bispectrumoriginal}
\end{equation}
We can rewrite Eq.~(\ref{bispectrumoriginal}) as
\begin{eqnarray}
 \left< \zeta(\vec{k_1}) \zeta(\vec{k_2}) \zeta(\vec{k_3}) \right> &\simeq& N_{,\phi_{\sigma}} N_{,\phi_{s}}^2 
 \left< Q_{\sigma}(\vec{k_1}) Q_{s}(\vec{k_2}) Q_{s}(\vec{k_3}) + (\mbox{perm}) \right> \nonumber\\
 &+& \frac{1}{2} N_{,\phi_{s}}^{2} N_{\phi_{s} \phi_{s}} \left< Q_{s}(\vec{k_1}) Q_{s}(\vec{k_2}) \left(Q_{s} \star Q_{s} \right)(\vec{k_3}) \right> + \rm{two \: \: perms}.
\label{bispectrumdeltan}
\end{eqnarray}
Here we used the fact that $N_{,\phi_{\sigma}}$ is much smaller than $N_{\phi_{s}}$ in Eq.~(\ref{ndecomposition}) and 
the dominant bispectrum of the fields is coming from the mixed adiabatic and entropy contributions. Note that the star $\star$ denotes the convolution and correlators higher than the four-point were neglected in the above equation.
From Eqs.~(\ref{bispectrumoriginal}) and (\ref{bispectrumdeltan}), we obtain
\begin{equation}
 B_{\zeta}(k_1, k_2, k_3) =  \left. \frac{1}{\cos^{2} \Theta} N_{,\phi_{\sigma}}^3 B_{Q_{\sigma}}(k_1, k_2, k_3) \right\rvert_{t = t_{i}} + 4 \pi^{4} P_{\mathcal{R}}^{2} \frac{\Sigma_{j}k_j^{3}}{\Pi_{j}k_j^{3}} \frac{N_{\phi_{s} \phi_{s}}}{N_{,\phi_{s}}^{2}},
\label{bispectrumresult}
\end{equation}
by using Eq.~(\ref{deltanentropic}) where the bispectrum of the scalar field perturbation $B_{Q_{\sigma}}(k_1, k_2, k_3)$ is defined as
\begin{equation}
 \left< Q_{\sigma}(\vec{k_1}) Q_{\sigma}(\vec{k_2}) Q_{\sigma}(\vec{k_3}) \right> = \left( 2 \pi \right)^{2} B_{Q_{\sigma}} (k_1, k_2, k_3) \delta^{(3)} (\vec{k_1}+\vec{k_2}+\vec{k_3}).
\end{equation}
Here we used the fact that the symmetrised mixed bispectrum has the same shape as the pure adiabatic bispectrum \cite{Langlois:2008wt} and $N_{,\phi_{\phi}}/N_{,\phi_{s}}
\sim \cos \Theta \ll 1$.
The non-linear parameter $f_{\rm{NL}}$ is define as \cite{Maldacena:2003}
\begin{equation}
f_{\rm{NL}} \equiv \frac{5}{6} \frac{\Pi_{j}k_j^{3}}{\Sigma_{j}k_j^{3}} \frac{B_{\zeta}}{4 \pi^{4} P_{\mathcal{R}} ^{2}}.
\label{overallfnl}
\end{equation}
If the non-Gaussianity is local, we can express the curvature perturbation as
\begin{equation}
 \zeta = \zeta_{\rm{n}} + \frac{3}{5} f_{\rm{NL}} \left(\zeta_{n}^{2} - \left<\zeta_{n}^{2} \right> \right),
\end{equation}
where $\zeta_{n}$ obeys Gaussian statistics. From Eqs.~(\ref{bispectrumresult}) and (\ref{overallfnl}), the nonlinear parameter $f_{\rm{NL}}$ is expressed as
\begin{equation}
 f_{\rm{NL}} = \left. \cos^{2} \Theta f_{NL}^{equil} \right\rvert_{\Theta=0}+ \frac{5}{6} \frac{N_{,\phi_{s} \phi_{s}}}{N_{,\phi_{s}}^{2}},
\label{fnlderived}
\end{equation}
where $\left. f_{NL}^{equil} \right\rvert_{\Theta=0}$ is the equilateral non-Gaussianity parameter in the single field model defined as
\begin{equation}
\left. f_{NL}^{equil} \right\rvert_{\Theta=0} = \frac{5}{6} \frac{\Pi_{j}k_j^{3}}{\Sigma_{j}k_j^{3}} \frac{B_{Q_{\sigma}}}{4 \pi^{4} N_{,\sigma} P_{Q_{\sigma}} ^{2}}
=-\frac{35}{108}\frac{1}{c_s^2}.
\end{equation}
The first term on the right hand side of Eq.~(\ref{fnlderived}),
\begin{equation}
f_{NL}^{equil} \equiv
\cos^{2} \Theta \left. f_{NL}^{equil} \right\rvert_{\Theta=0},
\end{equation}
comes from the bispectrum of the quantum fluctuation of the scalar fields generated under horizon scales and gives rise to the equilateral non-Gaussianity, Eq.~(\ref{fnlequilateral}). We can see that small $\cos \Theta$ suppresses $f_{NL}^{equil}$. In our model, the equilateral non-Gaussianity is obtained as 
\begin{equation}
f_{NL}^{equil} \sim  - 9.5.
\end{equation}
The second term on the right hand side of Eq.~(\ref{fnlderived}),
\begin{equation}
 f_{NL}^{local} \equiv \frac{5}{6} \frac{N_{,\phi_{s} \phi_{s}}}{N_{,\phi_{s}}^{2}},
\label{fnlfinallocal}
\end{equation}
is generated even if the field perturbations at the horizon crossing are Gaussian.
This is called local type non-Gaussianity. Numerically, we can compute the second derivative of $N$ with respect to $\phi_{s}$ as
\begin{equation}
 N_{,\phi_{s} \phi_{s}} = \frac{N\left(\phi_{\sigma}^{I}(t_{i}) + Q_{s} e_{s}^{I}  \right) - 2 N\left(\phi_{\sigma}^{I}(t_{i}) \right) + N\left(\phi_{\sigma}^{I}(t_{i}) - Q_{s} e_{s}^{I}  \right)}{Q_{s}^{2}},
\label{secondorderderivative}
\end{equation}
The result of our numerical computation shows that we have
\begin{equation}
 f_{NL}^{local} = \frac{5}{6} \frac{N_{,\phi_{s} \phi_{s}}}{N_{,\phi_{s}}^{2}} \sim 40.1
\end{equation}
in our model (see appendix \ref{numericalmethod} for the details of the numerical computations). Note that $f_{NL}^{local}$ becomes constant after the transition.
This is compatible with the current WMAP observation \cite{Komatsu:2010}
\begin{equation}
 -10 < f_{NL}^{local} < 74 
 \: \: \: \: \: \: \: \: \: \: \: \: \rm{at\:\: 95\% \:\: C.L.}
\end{equation}

\section{Summary and Discussions}
\label{summarydiscussions}
DBI inflation is the most plausible model that generates large equilateral non-Gaussianity. However, single field UV DBI models in string theory are strongly disfavoured by the current observations of the spectral index and equilateral non-Gaussianity. It has been shown that these constraints are significantly relaxed in multi-field models if there is a large conversion of the entropy perturbations into the curvature perturbation. In this paper, for the first time, we quantified this conversion during inflation in a model with a potential where a waterfall phase transition connects two different radial trajectories (see Fig.~\ref{waterfall}). 
This type of potential appears in string theory where the angular directions 
become unstable in the warped conifold, which connects two extreme trajectories \cite{Chen:2010}. We demonstrated that all the observational constraints can be satisfied while obeying the bound on the tensor to scalar ratio imposed in string theory models. The large conversion also creates large local type non-Gaussianity in general. In our model, this is indeed the case and we expect that large equilateral non-Gaussianity is generally accompanied by large local non-Gaussainity in multi-field DBI model. This prediction can be tested precisely by upcoming data from the Planck satellite.

There are a number of extensions of our study. In this paper, we study a toy two-field model. It would be important to study directly the potentials obtained in string theory to confirm our results although it is a challenge to compute the potential when the sound speed is small. The curve in the trajectory in field space, which is essential to evade the strong constraints in string theory and responsible for large local non-Gaussianity, can be caused not only by the potential but also by non-trivial sound speeds \cite{Emery:2012}. This happens in a model with more than one throat for example where there appear multiple different sound speeds. It would be interesting to compare the two cases to see if one can distinguish between them observationally. Finally, it has been shown that the multi-field effects enhance the equilateral type trispectrum for a given $f_{NL}^{equil}$ \cite{Mizuno:2009cv} and its shape has been studied in detail \cite{Mizuno:2010by}. Moreover, it was shown that there appears a particular momentum dependent component whose amplitude is given by $f_{NL}^{local} f_{NL}^{equil}$ \cite{RenauxPetel:2009sj}. Thus the trispectrum will provide further tests of multi-field DBI inflation models.

\begin{acknowledgments}
We would like to thank Jon Emery, Gianmassimo Tasinato and David Wands for useful discussions. TK and KK are supported by the Leverhulme trust. KK is also supported by STFC grant ST/H002774/1, the ERC starting grant. SM is supported by Labex P2IO in Orsay. SM is also grateful to the ICG, Portsmouth, for their hospitality
when this work was initiated.
\end{acknowledgments}

\appendix

\section{Adiabatic and entropy perturbations}
\label{Adiabaticandentropic}
We summarise how the adiabatic and entropic bases are defined here. We consider the linear perturbations of the scalar fields defined as
\begin{equation}
 \phi^{I}(x,t)=\phi^{I}_{0}(t) + Q^{I}\left(x,t\right).
\end{equation}
We decompose the perturbations into the instantaneous adiabatic and
entropy perturbations where the adiabatic direction corresponds to the direction of the background field's evolution while the entropy directions are orthogonal to this \cite{Gordon:2001}. For this purpose, we introduce an orthogonal basis $e^{I}_{\rm{n}} (n = 1,2...N)$ in field space.
The orthonormal condition in general multi-field inflation is given by
\begin{equation}
 P_{,X^{I J}} e^{I}_{n} e^{J}_{m} = \delta_{n m},
\label{normalization}
\end{equation}
so that the gradient term $P_{,X^{I J}} \partial_{\rm{i}} Q^{I} \partial^{\rm{i}} Q^{J}$ is diagonalised when we use this basis. Here we assume that $P_{,X^{I J}}$ is
invertible and it can be used as a metric in the field space. The adiabatic vector is
\begin{equation}
 e_{1}^{I} = \frac{\dot{\phi}^{I}}{\sqrt{P_{,X^{J K}} \dot{\phi}^{J} \dot{\phi}^{K}}},
\end{equation}
which satisfies the normalization given by Eq.~(\ref{normalization}). The field perturbations are decomposed in this basis as
\begin{equation}
 Q^{I} = Q_{n} e^{I}_{n}.
\end{equation}
For multi-field DBI inflation, using the relation
\begin{equation}
 P_{,X^{I J}} \dot{\phi}^{I} \dot{\phi}^{J} = c_{s} G_{IJ} \dot{\phi}^{I} \dot{\phi}^{J} + \frac{1-c_{s}^{2}}{2 X c_{s}} G_{IK} G_{JL} \dot{\phi}^{I} \dot{\phi}^{K} \dot{\phi}^{J} \dot{\phi}^{L} = \frac{2 X}{c_{s}},
\end{equation}
we can show that the adiabatic vector is given by
\begin{equation}
 e_{1}^{I} = \frac{\sqrt{c_{s}}}{2X} \dot{\phi}^{I}.
\label{adiabaticvector}
\end{equation}
This implies that 
\begin{equation}
 G_{IJ} e_{1}^{I} e_{1}^{J} = c_{s},
\end{equation}
\begin{equation}
 P_{,X^{IJ}} = c_{s} G_{IJ} + \frac{1 - c_{s}^{2}}{c_{s}^{2}} G_{IK} G_{JL} e_{1}^{K} e_{1}^{L}.
\label{canonicaldecomposition}
\end{equation}
Substituting Eq.~(\ref{canonicaldecomposition}) into Eq.~(\ref{normalization}), with $m = 1$, we obtain
\begin{equation}
 G_{I J} e_{1}^{I} e_{n} ^{J} = c_{s} \delta_{n 1}.
\label{orthogonal}
\end{equation}
Substituting Eqs.~(\ref{canonicaldecomposition}) and (\ref{orthogonal}) into 
Eq.~(\ref{normalization}), we obtain
\begin{equation}
 G_{I J} e_{n}^{I} e_{m} ^{J} = \frac{1}{c_{s}} \delta_{m n} - \frac{1 - c_{s}^{2}}{c_{s}} \delta_{m 1} \delta_{n 1}.
\label{finalnormalization}
\end{equation}
For two-field models with $G_{I J} \left(\phi^{K} \right) = A_{I} \left(\phi^{K} \right) \delta_{I J}$ where $A_{\phi} = 1$ and $A_{\chi} = \phi^{2}$, from 
Eq.~(\ref{adiabaticvector}), the adiabatic vector is obtained as
\begin{equation}
 \left(e_{\sigma}^{\phi}, e_{\sigma}^{\chi} \right) = \left(\sqrt{c_{s}}\frac{\dot{\phi}}{\sqrt{\dot{\phi}^{2}+\phi^{2}\dot{\chi}^{2}}}, \sqrt{c_{s}}\frac{\dot{\chi}}{\sqrt{\dot{\phi}^{2}+\phi^{2}\dot{\chi}^{2}}} \right).
\end{equation}
From the orthogonal condition (\ref{finalnormalization}), the entropy vector $e_{s}^{I}$ satisfies
\begin{equation}
 G_{IJ}e_{\sigma}^{I}e_{\sigma}^{J} = e_{\sigma}^{\phi}e_{s}^{\chi} + \phi^{2}e_{\sigma}^{\chi}e_{s}^{\chi} = 0,
\end{equation}
\begin{equation}
 G_{IJ}e_{s}^{I}e_{s}^{J} = \left(e_{s}^{\phi}\right)^{2} + \phi^{2}\left(e_{s}^{\chi}\right)^{2} = \frac{1}{c_{s}},
\end{equation}
which leads to 
\begin{equation}
 \left(e_{s}^{\phi}, e_{s}^{\chi} \right) = \left(\frac{1}{\sqrt{c_{s}}}\frac{\phi \dot{\chi}}{\sqrt{\dot{\phi}^{2}+\phi^{2}\dot{\chi}^{2}}}, \frac{1}{\sqrt{c_{s}}}\frac{-\dot{\phi}}{\phi\sqrt{\dot{\phi}^{2}+\phi^{2}\dot{\chi}^{2}}} \right).
\end{equation}

\section{Numerical method}
\label{numericalmethod}
In this section, we explain how the $\delta$N-formalism is used in the numerical computations. In the single field case, we just need to perturb the initial conditions along the trajectory. In the numerical computations, it is easy to compute $\phi_{\sigma}(t_{i} + \delta t)$ because the trajectory with the perturbed initial conditions is the same as the one with the original initial conditions.

However, in the two-field case, the trajectory with the initial conditions peruturbed in the entropic direction is different
from the one with the original initial conditions as is shown in Fig.~\ref{entropyperturbation}. Although the perturbed initial values of the
fields are defined in Eq.~(\ref{entropicperturbedfield}), once we set the value of $Q_{s}$, we also need to know how to perturb the values of the time derivatives of the fields. This is because we need to set the values of all the phase space variables in order to solve the second order differential equations numerically. From 
Eq.~(\ref{entropicperturbedfield}), we obtain
\begin{equation}
 \dot{\tilde{\phi}}_{\sigma}^{I}(\tilde{t}_{i}) = \dot{\phi}_{\sigma}^{I}(t_{i}) + \dot{Q}_{s} e_{s}^{I} + Q_{s} \dot{e}_{s}^{I}.
\label{timederivativeofpert}
\end{equation}
From Eqs.~(\ref{basisone}) and (\ref{basistwo}), the derivatives of the entropy basis vector are obtained as 
\begin{equation}
 \dot{e}_{s}^{\phi} = e_{s}^{\phi} \left(-\frac{f_{,\phi} \dot{\phi}}{2 f} - \frac{s H}{2} - \frac{\ddot{\sigma}}{\dot{\sigma}} + \frac{\dot{q}(t)}{q(t)} \right),
\label{esphidotrelation}
\end{equation}
and
\begin{equation}
 \dot{e}_{s}^{\chi} = e_{s}^{\chi} \left(-\frac{f_{,\phi} \dot{\phi}}{2 f} - \frac{s H}{2} - \frac{\ddot{\sigma}}{\dot{\sigma}} - \frac{\dot{\phi}}{\phi} + \frac{\dot{p}(t)}{p(t)} \right),
\label{eschidotrelation}
\end{equation}
Here new variables $p(t)$ and $q(t)$ are defined as
\begin{equation}
  p(t) \equiv -\sqrt{f} \dot{\phi}, \: \: \: \: \: q(t) \equiv -\sqrt{f} \phi \dot{\chi},
\end{equation}
so that the sound speed is expressed as
\begin{equation}
 c_{s} = \sqrt{1 - p(t)^{2} - q(t)^{2}}.
\end{equation}
Given that we obtain the numerical values of $\dot{e}_{s}^{\phi}$ and $\dot{e}_{s}^{\chi}$ using Eqs.~(\ref{esphidotrelation}) and (\ref{eschidotrelation}), we now know how to perturb all the phase space variables ($\phi$, $\dot{\phi}$, $\chi$, $\dot{\chi}$) from Eq.~(\ref{timederivativeofpert}) if we know the value of $\dot{Q}_{s}$. Because we set the value of
$Q_{s}$, we can determine the value of $\dot{Q}_{s}$ if there is a relation between $Q_{s}$ and $\dot{Q}_{s}$. Actually, before the transition, it is possible to obtain the analytic solution for $v_{s}$ and hence the solution for $Q_{s}$ from 
Eq.~(\ref{relationofvandq}). As mentioned in section~\ref{linearperturbation},
the approximations $z''/z \simeq 2 / \tau^{2}$ and $\alpha'' / \alpha \simeq 2 / \tau^{2}$ hold because the slow-roll approximation holds and the coupling
$\xi$ is negligible before the transition. Therefore, Eq.~(\ref{equationofmotiontwo}) is approximated as
\begin{equation}
 v_{s}'' + \left(c_{s}^{2} k^{2} - 2 / \tau^{2} + a^{2} \mu_{s}^{2} \right)v_{s}  \simeq 0,
\end{equation}
which can be rewritten as
\begin{equation}
 \frac{d^{2} \tilde{v}_{s}}{d \tilde{\tau} ^{2}} + \frac{1}{\tilde{\tau}} \frac{d \tilde{v}_{s}}{d \tilde{\tau}} + \left(1 - \frac{9/4 - \mu_{s}^{2}/H^{2}}{\tilde{\tau}^{2}} \right)\tilde{v}_{s}  \simeq 0,
\label{besselequation}
\end{equation}
where
\begin{equation}
  \tilde{v}_{s} \equiv \frac{v_{s}}{\sqrt{- \tau}}, \: \: \: \: \: \tilde{\tau} \equiv -c_{s} k \tau.
\end{equation}
Note that we regard $c_{s}$ as a constant because the slow-roll parameter $s$ is much smaller than unity as we showed in Sec.~\ref{linearperturbation}.
Then, if we approximate $\mu_{s}$ to be a constant, we have the analytic solution for Eq.~(\ref{besselequation}) because it is the Bessel differential equation.
By choosing the Bunch-Davies vacuum initial condition, we obtain
\begin{equation}
 v_{s} = \frac{\sqrt{\pi}}{2} e^{i (\nu_{s} + 1/2) \pi/2} \left(- \tau \right)^{1/2} H_{\nu_{s}}^{(1)}\left(-c_{s} k \tau \right),
\end{equation}
where
\begin{equation}
 \nu_{s} = \sqrt{9/4 - \mu_{s}^{2}/H^{2}},
\end{equation}
and $H_{\nu_{s}}^{(1)}$ is the Hankel function of the first kind. In the super-horizon limit $-c_{s} k \tau << 1$, using the asymptotic form of the Hankel function
\begin{equation}
 H_{\nu_{s}}^{(1)}\left(-c_{s} k \tau \right) \sim - i \frac{\Gamma(\nu_{s})}{\pi} \left(\frac{2}{-c_{s} k \tau} \right)^{\nu_{s}},
\end{equation}
we have
\begin{equation}
 v_{s} = \frac{- i \sqrt{\pi}}{2} e^{i (\nu_{s} + 1/2) \pi/2} \left(- \tau \right)^{1/2} \frac{\Gamma(\nu_{s})}{\pi} \left(\frac{2}{-c_{s} k \tau} \right)^{\nu_{s}},
\end{equation}
and
\begin{equation}
 Q_{s} = \frac{- i \sqrt{\pi} c_{s} H}{2} e^{i (\nu_{s} + 1/2) \pi/2} \left(- \tau \right)^{3/2} \frac{\Gamma(\nu_{s})}{\pi} \left(\frac{2}{-c_{s} k \tau} \right)^{\nu_{s}},
\label{qssuperhorizon}
\end{equation}
where we used the relation $\tau \sim -1/aH$ during slow-roll inflation. From 
Eq.~(\ref{qssuperhorizon}), we obtain the first derivative of $Q_s$ in terms of 
$Q_s$ as 
\begin{equation}
 \dot{Q}_{s} = \left[s\left(1 - \nu_{s} \right) - \epsilon + \nu_{s} - \frac{3}{2} \right] H Q_{s}.
\label{qdot}
\end{equation}

Let us show some numerical results in the model introduced in section~\ref{twofield}. As is shown in Fig.~\ref{chisound}, the transition begins around $N \sim 20$.
We take the initial hypersurface around $N \sim 12$ where we can evaluate $P_{Q_{\sigma}}$ with Eq.~(\ref{pqsigmas}) because the trajectory is still 
effectively a single field one after sound horizon crossing.
On the initial hypersurface, we set the values of all the phase space variables as $\phi(t_{i}) \simeq 7.09, \dot{\phi}(t_{i}) \simeq 8.11 \times 10^{-11}, \chi(t_{i}) \simeq 5.06 \times 10^{-5},
\dot{\chi}(t_{i}) \simeq 3.42 \times 10^{-14}$. If we set $\delta \chi = 10^{-7}$, we obtain
\begin{equation}
 \delta \phi = \frac{e^{\phi}_{s}(t_{i})}{e^{\chi}_{s}(t_{i})} \delta \chi = -2.12 \times 10^{-9},
\label{deltaphi}
\end{equation}
from Eq.~(\ref{entropicperturbedfield}) where $e^{\phi}_{s}(t_{i})$ and $e^{\chi}_{s}(t_{i})$ are obtained numerically. Then, we also have
\begin{equation}
 Q_{s} = \frac{10^{-7}}{e^{\chi}_{s}(t_{i})} = -5.59 \times 10^{-8}.
\label{qsnumerical}
\end{equation}
Using Eq.~(\ref{qdot}), we obtain $\dot{Q}_{s} \simeq -4.29 \times 10^{-17}$. Actually, we can also obtain the first derivative numerically
\begin{equation}
 \dot{Q}_{s} = \left( \frac{\dot{v_{s}}}{v_{s}} + H \left(s - 1 \right) \right) Q_{s} \simeq -3.07 \times 10^{-17},
\label{qsdotnumerical}
\end{equation}
from Eq.~(\ref{relationofvandq}) using the numerical values of $\dot{v}_{s}$ and $v_{s}$. We can see that the values of $\dot{Q}_{s}$ obtained in both ways are the same with around 40 percent error. This error comes from the fact that the solution 
Eq.~(\ref{qdot})is exact only if $\mu_{s}$ is perfectly constant and the slow-roll parameters are zero. Using Eqs.~(\ref{timederivativeofpert}), (\ref{qsnumerical}), (\ref{qsdotnumerical}) and the numerical values of $e_{s}^{I}$ and $\dot{e}_{s}^{I}$ obtained by using Eqs.~(\ref{basisone}), (\ref{basistwo}), (\ref{esphidotrelation}) and (\ref{eschidotrelation}), we obtain the first derivatives of the fields as 
\begin{equation}
 \delta \dot{\phi} \equiv \dot{\tilde{\phi}}_{\sigma}(\tilde{t}_{i}) - \dot{\phi}_{\sigma}(t_{i}) \simeq -1.47 \times 10^{-17},
\label{deltaphidot}
\end{equation}
and
\begin{equation}
 \delta \dot{\chi} \equiv \dot{\tilde{\chi}}_{\sigma}(\tilde{t}_{i}) - \dot{\chi}_{\sigma}(t_{i}) \simeq 5.35 \times 10^{-17}.
\label{deltachidot}
\end{equation}
Using $\delta \chi = 10^{-7}$, Eqs.~(\ref{deltaphi}), (\ref{deltaphidot}) and (\ref{deltachidot}), we can now perturb the initial conditions. 
Using these initial condition, the first derivative of $N$ with respect to $\phi_s$
is obtained using Eqs.~(\ref{deltanentropy}) and (\ref{nphis}) as
\begin{equation}
N_{,\phi_{s}} \simeq 7.82 \times 10^{2}.
\end{equation}
Then, from Eq. (\ref{deltanentropic}), the power spectrum of the curvature 
perturbation is given by 
\begin{equation}
 P_{\mathcal{R}} = P_{\zeta} \simeq N_{,\phi_{s}}^{2} P_{Q_{s}} \rvert _{t=t_{i}} \simeq 2.29 \times 10^{-9},
\end{equation}
where we used the numerical result for the power spectrum of the entropy perturbation
\begin{equation}
 P_{Q_{s}} \rvert _{t=t_{i}} = \frac{k^{3}}{2 \pi^{2}} \frac{c_{s}^{2} \lvert v_{s k} \rvert^{2}}{a^{2}} \simeq 3.74 \times 10^{-15}.
\end{equation}
This coincides with the value $P_{\mathcal{R}} \sim 2.3 \times 10^{-9}$ with less than one percent error, which is obtained directly by solving the equations of motion for the linear perturbations numerically. This confirms the validity of the $\delta$N formalism in this model. In a similar way, we perturb $N$ and obtain the second order derivative using Eq.~(\ref{secondorderderivative}). Then it is possible to compute the local type non-Gaussianity.


\end{document}